\documentclass[prb,twocolumn,amsmath,amssymb,showpacs]{revtex4-1} 
\pdfoutput=1
\usepackage{graphics}
\usepackage{dcolumn}
\usepackage{bm}
\usepackage{color}
\usepackage{epstopdf}
\usepackage{epsfig}
\usepackage[colorlinks,linkcolor=blue,citecolor=magenta]{hyperref}
\begin{document}
\newcommand{\ket}[1]{\left|#1\right>}
\newcommand{\bra}[1]{\left<#1\right|}
\newcommand{\hlt}[1]{\textcolor{red}{#1}}
\title{Competing superconducting phases in interacting two-dimensional electron gas with strong Rashba spin-orbit coupling}
	\author{Rasoul Ghadimi}
	\affiliation{Department of Physics, Sharif University of Technology, Tehran 14588-89694, Iran}
	\author{Mehdi Kargarian}
	\email{kargarian@physics.sharif.edu}
	\affiliation{Department of Physics, Sharif University of Technology, Tehran 14588-89694, Iran}
	\author{S. Akbar  Jafari}
	\affiliation{Department of Physics, Sharif University of Technology, Tehran 14588-89694, Iran}
	
	\date{\today}
	\begin{abstract}

In this work we study interacting electrons on square lattice in the presence of strong Rashba spin-orbit interaction. The spin-orbit term forces the time-reversal electron states to be paired in even Cooper channels. For concreteness, we only consider the repulsive onsite Hubbard and nearest-neighbor coulomb interactions, the so called extended Hubbard model. To examine the superconducting instability we obtain the effective interaction between electrons within the random phase approximation and treat the pairing instabilities driven by charge and spin fluctuations and their combined effects. We mapped out the phase diagram of the model in terms of interactions and electron fillings, and found that while the $d_{xy}$ and $d_{x^2-y^2}$ symmetries are the most likely pairing symmetries driven by charge and spin fluctuations, respectively, the strong effect of both fluctuations yields higher angular momentum Cooper instability. The possibility of topological superconductivity and triplet pairing is also discussed.  
	\end{abstract}
	\maketitle
\section{Introduction}
Despite the ongoing tremendous efforts in the past decades to understand the unconventional superconductors, the pairing mechanisms and symmetry of the paired states continue to be important questions and, yet, in many cases remained to be unknown. In most cases the experimental evidences point to the  existence of nontrivial pairings not caused by phonons, giving rise to complicated structures for the gap function~\cite{mineev1999introduction,sigrist2005introduction,PhysRevB.39.11663}. For instance, the electronic spin density fluctuations may develop Cooper pairs with higher angular momentum such as $d$-wave, as opposed to fully symmetric and isotropic $s$-wave pairings~\cite{RevModPhys.84.1383,scalapino1986d}, e.g. in the high-$T_{c}$ superconductors as prime examples of unconventional superconductivity. 

Beside the pairing mechanisms, the spatial crystal symmetry may also influence the symmetry of the gap wave functions. In noncentrosymmetric 
superconductors~\cite{bauer2012non,yokoyama2007enhanced,samokhin2015symmetry}, due to the lack of inversion symmetry in the bulk of the underlying crystal, pairing states with mixed parities are expected to constitute the condensate. For example in CePt$_3$Si, the $s+p$-wave Cooper pairs may be realized~\cite{yokoyama2007enhanced},  though in this particular case a more careful study of the phase of the Cooper pairs by Samokhin, {\em et. al.}, indicates that the  order parameter in this system is an odd function of momentum 
that supports line of zero energy modes in the excitation spectrum~\cite{PhysRevB.69.094514}.

The two-dimensional electron gas (2DEG) confined at the interfaces between two insulators, which is the focus of this work, may also become a superconductor at low temperatures. One famous example of such 2DEG is the interface of LaAlO$_3|$SrTiO$_3$ system \cite{he2014two,PhysRevLett.115.147003,biscaras2010two,richter2013interface,PhysRevB.86.125121,PhysRevB87014510,PhysRevB.97.245113}, where the interfacial superconductivity offers an interesting playground for realizing the unconventional superconductivity~\cite{Reyren1196,yada2009electrically,nakamura2013multi,nakosai2012topological,GARIGLIO2015189,mannhart_blank_hwang_millis_triscone_2008,ohtomo2004high,saito2017highly}. At the interface the inversion and mirror symmetries are broken and consequently an interfacial Rashba spin-orbit coupling emerges giving rise to mixed singlet-triplet and multi-orbital superconductivities~\cite{gor2001superconducting, scheurer2015topological, samokhin2015symmetry}. In the presence of spin-orbit interaction the Cooper pairs acquire more robustness against dephasing in the magnetic fields beyond the Pauli paramagnetic limit~\cite{PhysRevLett.92.097001,takimoto2012mechanism,bauer2005superconductivity,0953-2048-29-12-123001,nakamura2013multi,1742-6596-807-5-052013}. Perhaps, the main advantage of studying superconductivity in heterostructures relies on its tunability by charge carriers or electric field \cite{PhysRevLett.108.247004,Ye1193,PhysRevLett.94.197004,ueno2008electric}. The spin-orbit coupling can be externally induced to tune the critical temperature of the system~\cite{PhysRevMaterials.2.024801,PhysRevLett.104.126803}. In the proximity to a conventional superconductor and in the presence of a Zeeman coupling or a magnetic field, the spin-orbit coupled 2DEG may host a topological superconductor~\cite{PhysRevLett.104.040502,PhysRevB.82.214509}. The emergence of the latter in heterostructures made of stacked 2DEGs with Rashba spin-orbit coupling 
has been studied theoretically~\cite{PhysRevLett.108.147003}. In the presence of strong disorder, a finite-momentum paired state, the so-called   Fulde-Ferrell-Larkin-Ovchinnikov (FFLO) state can be established~\cite{michaeli2012superconducting}. The in-plane magnetic field can also generate an FFLO state~\cite{PhysRevB.93.214516,larkin1965sov,PhysRev135A550}. While the out-of-plane magnetic fields may establish a $p+ip$ superconductor, for in-plane fields a nodal p-wave pairing is predicted~\cite{PhysRevB.93.214516,hugdal2018p}. Further, the in-plane field can induce a supercurrent~\cite{yip2002two}. As a function of charge carrier, the behavior of the $T_c$ in the interface of LaAlO$_3|$SrTiO$_3$ is found to be nonmonotonic~\cite{PhysRevB.89.184514}, and the pairing symmetry in this system can be controlled by an applied electric field~\cite{yada2009electrically}. 

Hence, the ability to tune the parameters of 2DEGs, which are by now within the experimental reach and controllability, provide a fertile ground enabling us to study the interplay between strong correlations and spin-orbit interaction. It is shown that the strong Hubbard interaction and spin-orbit coupling give rise not only to superconducting instabilities \cite{vafek2011spin, doi:10.7566/JPSJ.82.014702}, but also to time-reversal symmetry-breaking superconducting states with even angular momenta~\cite{vafek2011spin} and topological superconductivity~\cite{scheurer2015topological}. In the absence of spin-orbit coupling the Hubbard model treated within the random phase approximation (RPA) on a square lattice yields a chiral p-wave superconducting state that breaks the time-reversal symmetry~\cite{romer2015pairing}. It is attributed to the enhancement of the spin susceptibility near $\mathbf q=(0,0)$.  
A phase transition from the $p$-wave state at very low filling to $d_{x^2-y^2}$-wave symmetry 
close to half-filling is also reported. The same system including the Rashba spin-orbit interaction has been studied in Ref.~[\onlinecite{greco2017topological}]. It's shown that the ferromagnetic fluctuations are dominant for values of chemical potentials lying between the van-Hove singularities resulting in a possible $f$-wave triplet pairing.
	
In this work we consider the {\em extended} Hubbard model, including both onsite and nearest-neighbor Coulomb interactions, on the square lattice in the presence of strong Rashba spin-orbit coupling. Our goal is to envisage the role of strong correlations, spin-rotational symmetry-breaking effects, and the electron fillings on the formation of the Cooper pairs. In particular, we (i) derive the effective interaction between electrons dressed by spin and charge fluctuations within the RPA, (ii) use the even-parity pairing states between time-reversed states on the Fermi contours, dictated by strong spin-orbit coupling, to investigate the superconducting instability, (iii) obtain the superconducting phase diagram and the phase transition between different superconducting states over a wide range of interactions and fillings, and (iv) discuss the origin of the triplet superconductivity in this system.  

The paper is organized as follows. We start by describing the model in section \ref{modelmethod}. In Sec.~\ref{effectivinteraction} we will derive the effective interaction between the electrons. The Sec.~\ref{pairingsymmetryinstability} is devoted to the pairing instabilities  and their symmetries. In Sec.~\ref{results} we present our results and finally we conclude in Sec.~\ref{conclusions}.

\section{interacting 2DEG Model}
		\label{modelmethod}
		 \begin{figure*}
		\centering
		\includegraphics[width=0.9\linewidth]{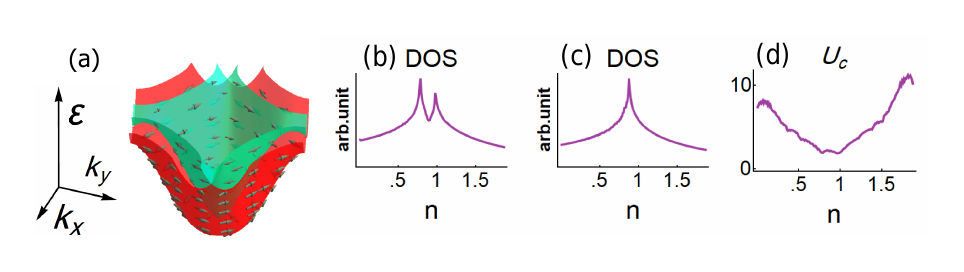}
		\caption{ (a) Energy dispersion of the noninteracting electrons. The color and arrows show different helicity and orientation of spins respectively. 
		Parameters are $t=1$ , $t'=0.3$ and $V_{so}=0.5$. Density of state at the Fermi level versus filling with (b) $(t,t',V_{so})=(1,0.3,0.5)$
		and (c) $(t,t',V_{so})=(1,0.3,0)$. (d) Critical interaction, where spin susceptibility diverges, as a function of filling $n$ for $U_1=0.0$. }
		\label{fighelicity}
	\end{figure*}
The model we consider consists of a kinetic term $H_{0}$ and an interaction $H_{I}$ between electrons, $H=H_{0}+H_{I}$. We assume that the noninteracting electrons on a square lattice are described by a single-particle Hamiltonian 
	\begin{equation}
	\label{ham}
	H_0=\sum_{\mathbf{k}} \psi^\dagger_{\mathbf{k}}\left(\varepsilon_{\mathbf k}\sigma_0+\mathbf g_{\mathbf{k}}\cdot\boldsymbol \sigma\right)\psi_{\mathbf{k}},
	\end{equation}
	where $\psi^T_{\mathbf{k}}=\left(\begin{matrix}
	c_{\uparrow{\mathbf{k}}},c_{\downarrow{\mathbf{k}}}
	\end{matrix}\right)$ with $c_{\sigma{\mathbf{k}}}$($c^\dagger_{\tau{\mathbf{k}}}$) as the annihilation(creation)  of electron with spin $\tau$  and momentum $\mathbf{{\mathbf{k}}}$.
In the Hamiltonian (\ref{ham}) $\varepsilon_{\mathbf k}=-2t(\cos  k_xa+\cos  k_ya)+t'\cos  k_xa\cos  k_ya-\varepsilon_F$ is the 2D energy dispersion in the absence of spin-orbit interaction, where   $t,t',\varepsilon_F$ are the nearest, next-nearest hopping amplitudes, and the Fermi energy, respectively. In the following we set lattice constant to be unity $a=1$.
The spin-rotational and inversion symmetries are broken by adding the spin-orbit Rashba interaction 
 $\mathbf g_{\mathbf{k}}=V_{so} \mathbf\nabla \varepsilon_{\mathbf k}\times\hat z$ with strength $V_{so}$, where $\hat z$ is  unit vector  perpendicular to the interface. Also $\sigma_0$ is identity matrix with dimension two and $\boldsymbol{\sigma}=(\sigma^x,\sigma^y,\sigma^z)$ is a vector of Pauli matrices.
 
The Hamiltonian (\ref{ham}) can be diagonalized by introducing band basis creation and annihilation operators as
\begin{equation}
a^\dagger_{\mathbf{k},\lambda}=\frac{1}{\sqrt{2}}\left(
i	\lambda e^{i\phi(\mathbf{k})}c_{{\mathbf{k}},\uparrow}^\dagger+c_{{\mathbf{k}},\downarrow}^\dagger\right),\quad  ie^{i\phi_{\mathbf{k}}}=\frac{g_{x\mathbf{k}}-i g_{y\mathbf{k}}}{|g_{\mathbf{k}}|}.
\label{acreation}
\end{equation}
It then follows that
\begin{equation}
H_0=\sum_{{\mathbf{k}}\lambda}\varepsilon_{\lambda\mathbf{k}} a^\dagger_{\mathbf{k}\lambda}a_{\mathbf{k}\lambda},
\end{equation}
where $\lambda=\pm 1$ label non-degenerate bands with dispersion
\begin{equation}
\label{dispersion}
\varepsilon_{\lambda\mathbf{k}}=\varepsilon_{\mathbf k}+\lambda |\mathbf g_{\mathbf{k}}|.
\end{equation}
In Fig. (\ref{fighelicity}a) we show the energy dispersion of the bands for a given set of parameters. It is clearly seen that the $\mathbf g_{\mathbf k}$ lift the spin degeneracy resulting in two non-degenerate bands throughout the Brillouin zone except for a few exceptional points, the so-called Kramers' degeneracy at the time-reversal invariant momenta, where $\mathbf g_{\mathbf k}=0$. Also, one notes that  the spin is locked to the momentum as shown by arrows. Given a fixed generic value for Fermi energy $\varepsilon_F$, two separate Fermi contours can be distinguished with opposite helicity $\lambda$ define by $\widehat{\mathbf g}(\mathbf k).\boldsymbol{\sigma}\left|\mathbf k\lambda\right> =\lambda \left|\mathbf k\lambda\right> $, where $\left|\mathbf k\lambda\right> $ is eigenvector of the Hamiltonian and hat denotes the unit vector.

For describing the repulsive interaction between electrons we use on site Hubbard and nearest-neighbor repulsive interactions  
	\begin{eqnarray}
	\label{hubbard}
H_I=U\sum_{i}n_{i\uparrow}n_{i\downarrow}+\frac{U_1}{2}\sum_{	<ij>,
	\tau\tau'}n_{i\tau}n_{j\tau'},
	\end{eqnarray}
where	$n_{i\tau}=c^\dagger_{i\tau}c_{i\tau}$ is the electron occupation number operator with spin $\tau$ at site $i$. Here $U$ and $U_1$ are the strength of Hubbard and nearest-neighbor interactions, respectively. In the next section, we obtain an effective interaction between electrons within the random phase approximation (RPA) before turning to the Cooper instability of the Fermi contours in the following sections.

	\section{Effective Interaction}
	\label{effectivinteraction}
In this section we derive effective interaction between electrons using  (RPA). To begin, we rewrite the Hubbard interaction (\ref{hubbard}) in the momentum space (see appendix \ref{Interactionmatrix} for details of derivation)
	\begin{equation}
	\label{HI}
	H_I=\frac{1}{N}\sum_{\mathbf{q},\alpha,\beta} \rho_{\alpha,\mathbf{q}} V_{\alpha\beta}\rho_{\beta,-\mathbf{q}},
	\end{equation}
	where $\alpha,\beta\in\{0,x,y,z\}$  and
	\begin{equation}
	\label{effinteraction}
			\hat{V}(\mathbf q)=\left(\begin{matrix}
		U_0(\mathbf q)&0&0&0\\
		0&-U&0&0\\
		0&0&-U&0\\
		0&0&0&-U
		\end{matrix}\right),
	\end{equation}
	where $U_0(\mathbf q)=U+U_1\left(\cos q_x+\cos q_y\right)$
and $N$ is the total number of sites. The charge ($\alpha=0$) and the spin ($\alpha=\{x,y,z\}$) density operators are expressed  as
	\begin{equation}
	\label{density}
	\rho_{\alpha,\mathbf q}=\sum_{\mathbf k\tau\tau'} c^\dagger_{\mathbf k+\mathbf q,\tau}\sigma^{\alpha}_{\tau,\tau'} c_{\mathbf k,\tau'}.
	\end{equation} 
Within the RPA,  the effective interaction is given by
	\begin{equation}
	\label{Veff}
	\hat{V}^{\rm eff}(\omega,\mathbf q)=\frac{\mathbf 1}{\mathbf 1- \hat{V}(\mathbf{q}) \hat{\chi}^R(\omega,\mathbf q)}\hat{V}(\mathbf q),
	\end{equation}
	where $\omega$ and $\mathbf q$ denote the frequency and momentum, respectively, and $\chi^R$ is the retarded charge and spin density susceptibility matrix
\begin{small}
		\begin{equation}
	\label{XR}
	\hat{\chi}^R(\omega,\mathbf q)=\sum_{\lambda \lambda'=\pm 1}\int \frac{d^2k}{4\pi^2} \frac{n_F(\varepsilon_{\lambda\mathbf k})-n_F(\varepsilon_{\lambda'\mathbf k+\mathbf q})}{\omega+i0^++\varepsilon_{\lambda\mathbf k}-\varepsilon_{\lambda'\mathbf k+\mathbf q}}\hat{F}_{\mathbf k,\mathbf k+\mathbf q;\lambda,\lambda'}.
	\end{equation}
\end{small}

	The Fermi Dirac distribution is give by $n_F(\epsilon)$ and form factor matrix $F$ is\cite{pletyukhov2007charge}
	\begin{equation}
	\label{formfactor}
	F^{\alpha\beta}_{\mathbf k,\mathbf k+\mathbf q;\lambda,\lambda'}={\rm tr}(\sigma^{\beta}\hat P_{\mathbf k\lambda}\sigma^{\alpha}\hat P_{\mathbf k+\mathbf q\lambda'}),
	\end{equation}
with the projection operator $\hat P$  defined as $\hat P_{k,\lambda}=\left|\mathbf k\lambda\right>\left<\mathbf k\lambda\right|$. 
In the static limit, $\omega\rightarrow 0$, the hermiticity of the interaction implies that the spin and charge components of the susceptibilities, and consequently effective interaction, decouple from each other (see appendix \ref{vanishing} for more details). In the spin (charge) channel at a critical value of $U_c$ ($U_{1c}$) the determinant of the denominator of (\ref{Veff}) vanishes, $\det\left(\mathbf{1}- \hat{V}(\mathbf{q}) \hat{\chi}^R(\omega,\mathbf q)\right)=0$, implying an instability of the system to spin-density wave (charge-density wave) state denoted by SDW (CDW).\\

	\section{Superconductivity and PAIRING SYMMETRY}
	\label{pairingsymmetryinstability}
	\subsection{BCS-like superconductivity and even-parity condensate}
	\label{BCSLIKESUPER}
  In this work, we only analyze the case of Bardeen-Cooper-Schrieffer (BCS) - like superconductivity which means that the Cooper pairs have a vanishing center of mass momentum.
BCS pairing occurs between two electron states with opposite momenta residing on the Fermi contours. Since the Fermi contours are single-degenerate due to the spin-rotational symmetry breaking nature of the Rashba coupling, it is more convenient to rewrite the interaction in the band basis $\left|\mathbf k\lambda\right>$\cite{samokhin2008gap,samokhin2015symmetry}.  

Inverting  Eq. (\ref{acreation}), we can express the electron operators $c^{\dagger}_{\mathbf k\tau}$ in terms of band operators $a^{\dagger}_{\mathbf k\lambda}$ as
	\begin{equation}
	\label{invertion}
	c_{{\mathbf{k}}\tau}^\dagger=\frac{1}{\sqrt{2}}\sum_{\lambda}(-t_{\lambda\mathbf{k}})^{\frac{\tau+1}{2}}a^{\dagger}_{{\mathbf{k}},\lambda},
	\end{equation}
	where $t_{\lambda\mathbf{k}}=i\lambda e^{-i\phi_{\mathbf{k}}}$ which is odd under $\mathbf k\rightarrow -\mathbf k$ since $\phi_{-\mathbf k}\rightarrow\phi_{\mathbf k}+\pi$. Note that in writing (\ref{invertion}), in the sum we use $\tau=+1$ ($-1$)  for spin up (down).
	
In following, because we  are only interested in finding the paring symmetry, we just consider the static limit of the effective interaction $\hat{V}(\mathbf{q})=\hat{V}^{{\rm eff}}(0,\mathbf{q})$. We use (\ref{invertion})  to rewrite the effective interaction
		\begin{equation}
	\label{HIg}
	H_I=\frac{1}{N}\sum_{\mathbf{q},\alpha} \rho_{\alpha,\mathbf{q}} V_{\alpha\beta}(\mathbf q)\rho_{\beta,-\mathbf{q}},
	\end{equation}
	in terms of band  basis operators as
\begin{widetext}
	\begin{equation} \label{HI_band}
		H_I=\frac{1}{N}\sum_{\{\lambda_{i}\},\{\tau_{i}\}, \alpha,\beta}\sum_{\mathbf{k}, \mathbf{q}} \sigma^{\alpha}_{\tau_1,\tau_2}V_{\alpha\beta}(\mathbf q)\sigma^{\beta}_{\tau_3,\tau_4}
	\frac{(-t_{\lambda_1{\mathbf{k}}+\mathbf q})^{\frac{\tau_1+1}{2}}(-t^{*}_{\lambda_2\mathbf{k}})^{\frac{\tau_2+1}{2}}}{(-t^*_{\lambda_3{\mathbf{k}}'})^{\frac{\tau_3+1}{2}}(-t_{\lambda_4{\mathbf{k}}'+\mathbf q})^{\frac{\tau_4+1}{2}}}	 a^\dagger_{{\mathbf{k}}+\mathbf q,\lambda_1}a_{{\mathbf{k}},\lambda_2}a^\dagger_{{\mathbf{k}}',\lambda_3}a_{{\mathbf{k}}'+\mathbf q,\lambda_4}.
	\end{equation}
\end{widetext} 

Since we are interested in the  pairing between electrons with opposite momenta, we restrict the momentum summation in (\ref{HI_band}) to the Cooper channel. Moreover, we consider the pairing between an electron in the state $\left|\mathbf k\lambda\right>$ and its time-reversal partner $\left|\tilde{\mathbf  k}\lambda\right>=\Theta \left|\mathbf k\lambda\right>=t_{\lambda\mathbf k}\left|-\mathbf k\lambda\right>$, where $\Theta=i\sigma_y \mathcal K$ with $\mathcal K$ as complex conjugate operator\cite{Gonge1602579,samokhin2008gap}. Consequently, the corresponding electron operators are related to each other by the following relations:
	\begin{equation}
	\begin{matrix}
	\tilde{a}_{{\mathbf{k}}\lambda}^\dagger=\Theta a^\dagger_{{\mathbf{k}},\lambda}\Theta^{-1}=t_{\lambda\mathbf{k}}a^\dagger_{-{\mathbf{k}},\lambda},\\
	\tilde{a}_{{\mathbf{k}}\lambda}=\Theta a_{{\mathbf{k}},\lambda}\Theta^{-1}=t^*_{\lambda\mathbf{k}}a_{-{\mathbf{k}},\lambda}.\\
	\end{matrix}
	\end{equation}
	Hence the effective interaction in the Cooper channel becomes\cite{samokhin2008gap}
	\begin{eqnarray}
	\label{interactionterm}
		H_I=\frac{U}{N}\sum_{{\mathbf{k}} {\mathbf{k}}'\lambda\lambda'}V_{\lambda',\lambda}({\mathbf{k}}',{\mathbf{k}}) a^\dagger_{{\mathbf{k}}',\lambda'}\tilde a^\dagger_{{\mathbf{k}}',\lambda'}\tilde a_{{\mathbf{k}},\lambda}a_{{\mathbf{k}},\lambda},
	\end{eqnarray}
	where
	\begin{eqnarray}
	\label{Veffectivetr}
	V_{\lambda',\lambda}({\mathbf{k}}',{\mathbf{k}})=-\sum_{\{\tau_i\},\alpha,\beta}
	\frac{\sigma^{\alpha}_{\tau_1,\tau_2}V_{\alpha\beta}({\mathbf{k}}'-{\mathbf{k}})\sigma^{\beta}_{\tau_3,\tau_4}
	(-1)^{\frac{\tau_1+\tau_2}{2}}}{(t^{*}_{\lambda'\mathbf{k}'})^{\frac{\tau_1+\tau_3}{2}}(t_{\lambda\mathbf{k}})^{\frac{\tau_2+\tau_4}{2}}}.\nonumber\\
	\end{eqnarray}
	Note that, because of decoupling of the charge and spin  susceptibilities we can decompose Eq. (\ref{Veffectivetr}) into the charge and spin sectors as
	\begin{equation}
		V_{\lambda',\lambda}({\mathbf{k}}',{\mathbf{k}})=V^{\rm charge}_{\lambda',\lambda}({\mathbf{k}}',{\mathbf{k}})+V^{\rm spin}_{\lambda',\lambda}({\mathbf{k}}',{\mathbf{k}})
	\end{equation}
	by restricting the summation in Eq. (\ref{Veffectivetr}) to $\alpha,\beta=0$ ($\alpha,\beta=1,2,3$) for charge (spin) sectors of effective interaction. 
	
By inspection we can see that in the above interaction, due to the relations $V_{\lambda',\lambda}({\mathbf{k}}',{\mathbf{k}})=V_{\lambda',\lambda}(- {\mathbf{k}}',{ \mathbf{k}})=V_{\lambda',\lambda}({\mathbf{k}}',{ -\mathbf{k}})$, only the {\em even} channels of $V_{\lambda',\lambda}({\mathbf{k}}',{\mathbf{k}})$,  survive. The reason is as follows. The Cooper pair annihilation operator $\hat b_{\lambda\mathbf{k}}=\tilde a_{{\mathbf{k}},\lambda}a_{{\mathbf{k}},\lambda}$ is an even function of momentum\cite{samokhin2008gap}, since 
 $\hat b_{\lambda\mathbf{k}}=
 t^*_{\lambda\mathbf{k}} a_{{-\mathbf{k}},\lambda} t_{\lambda-\mathbf{k}} \tilde a_{-{\mathbf{k}},\lambda}=\hat b_{\lambda\mathbf{-k}}$. Consequently, we can decompose the interaction $V$ in terms of only even basis functions of the irreducible representations of the point group symmetry of the underlying lattice. That is\cite{samokhin2008gap}
 \begin{equation}
 \label{basis}
 	V^s_{\lambda',\lambda}({\mathbf{k}}',{\mathbf{k}})=\sum_{a} V_{\lambda',\lambda}^{a}\sum_{i=1}^{d_a} \phi_{a,i}(\mathbf k')\phi^*_{a,i}(\mathbf k),
 \end{equation}
 where $a$ labels $d_a$-dimensional irreducible representations and 	 $V^s_{\lambda',\lambda}({\mathbf{k}}',{\mathbf{k}})=\frac{1}{2}\left(V^{\rm eff}_{\lambda',\lambda}({-\mathbf{k}}',{\mathbf{k}})+V^{\rm eff}_{\lambda',\lambda}({\mathbf{k}}',{-\mathbf{k}})\right)$ is symmetric part of the interaction. Here $\phi_{a,i}(\mathbf k)$ are even basis functions. 
 
 Using the mean field theory we decompose the interaction  (\ref{interactionterm}) as
	\begin{equation}
	\label{HIMF}
	H_{MF}=\sum_{{\mathbf{k}}\lambda}\left( \Delta_{\lambda}(\mathbf{k}) a^\dagger_{{\mathbf{k}}',\lambda'}\tilde a^\dagger_{{\mathbf{k}}',\lambda'}+{\rm h.c.}\right),
	\end{equation}
	where we ignored an unimportant constant term and the
 gap function	$\Delta_{\lambda}(\mathbf{k})=\Delta_{\lambda}(-\mathbf{k})$  is   given by
	\begin{equation}
	\label{gapfuncton}
	\Delta_{\lambda}(\mathbf{k})=\frac{1}{N}\sum_{{\mathbf{k}}'\lambda'} V^s_{\lambda,\lambda'}({\mathbf{k}},{\mathbf{k}}')  \langle \hat{b}_{\lambda'{\mathbf{k}}'}\rangle .
	\end{equation}
	
	 Including the single particle Hamiltonian $H_0$, the full BCS Hamiltonian becomes
	\begin{equation}
	\label{BCS}
	H_{BCS}=\sum_{{\mathbf{k}}\lambda}\left(\begin{matrix}
	a^\dagger_{{\mathbf{k}}\lambda}&\tilde a_{{\mathbf{k}}\lambda}
	\end{matrix}\right)
	\left(\begin{matrix}
	\epsilon_{\lambda\mathbf{k}}&\Delta_{\lambda}(\mathbf{k})\\\Delta^*_{\lambda}(\mathbf{k})&-\epsilon_{\lambda\mathbf{k}}
	\end{matrix}\right)
	\left(\begin{matrix}
	a_{{\mathbf{k}}\lambda}\\\tilde a^\dagger_{{\mathbf{k}}\lambda}
	\end{matrix}\right).
	\end{equation}
Using the Bogoliubov transformation, the energy dispersion of quasi-particle reads 
	\begin{equation}
	\label{qsiparticle}
	E_{\lambda\mathbf{k}}=\sqrt{\epsilon^2_{\lambda\mathbf{k}}+|\Delta_{\lambda}(\mathbf{k})|^2}.
	\end{equation}
The gap function (\ref{gapfuncton}) can be determined  self consistently as follows:
	\begin{equation}
	\label{delta}
	\Delta_{\lambda}(\mathbf{k})=-\frac{1}{N}\sum_{{\mathbf{k}}'\lambda'} V_{\lambda,\lambda'}^s({\mathbf{k}},{\mathbf{k}}') \frac{\Delta_{\lambda'}(\mathbf{k}') \tanh(\frac{1}{2}\beta E_{\lambda'\mathbf{k}'})}{2E_{\lambda'\mathbf{k}'}} .
	\end{equation}
		
		\begin{figure*}[t]
		\includegraphics[width=1\linewidth]{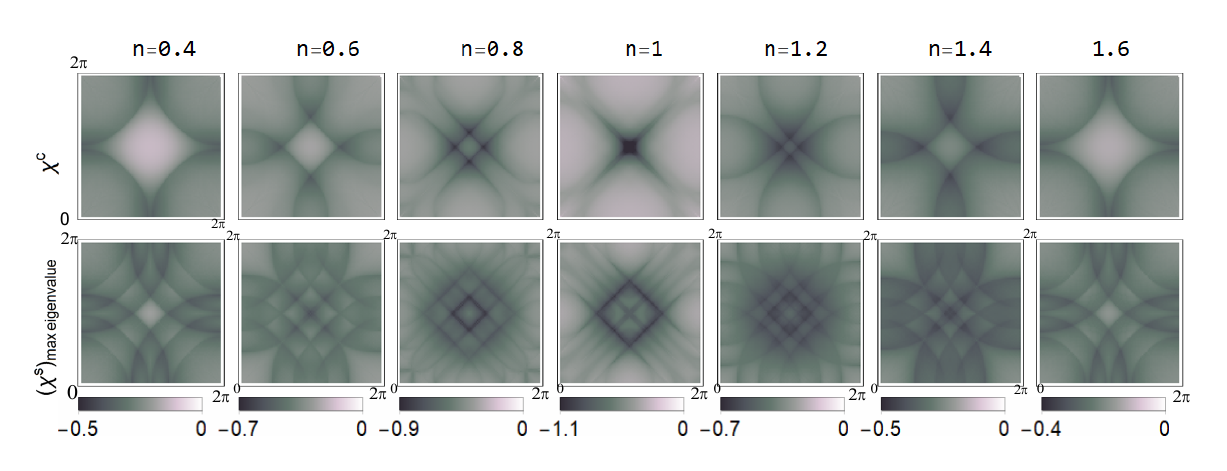}
		\caption[Fermi surface]{(Color online) Maximum eigenvalue of susceptibility matrix in the charge (upper panels) and spin (lower panels) channels at different fillings $n$.}
		\label{figmeshsucep}
	\end{figure*}

	 \subsection{Determination of the  gap function}
	 \label{DOGF}
	 Near the critical temperature  $T_c$, the gap equation can be linearized. Converting the sum in (\ref{delta}) to an energy integral about the Fermi contours with energy cut off $\omega_c$ and the momentum integration along the Fermi contours, we obtain\cite{romer2015pairing}
\begin{small}
		\begin{equation}
	\label{linergap}
	\Delta_{\lambda}(\mathbf{k})=-\ln\left(\frac{1.13\omega_c}{T_c}\right)\sum_{\lambda'}\int_{{FS}_{\lambda'}} \frac{dk'}{v_F^{\lambda'}(\mathbf{k}')} V_{\lambda,\lambda'}^s(\mathbf{k},\mathbf{k}')\Delta_{\lambda'}(\mathbf{k}') ,
	\end{equation}
\end{small}
where ${FS}_{\lambda}$ stands for  Fermi contour  and   $v^{\lambda}_F({\mathbf{k}})=|\nabla \varepsilon_{\lambda\mathbf{k}}|$ is the  $\mathbf k$-dependent Fermi velocity.

To obtain the pairing symmetry, we convert the gap equation (\ref{linergap}) to an eigenvalue problem by inserting Eq. (\ref{basis}) into (\ref{linergap})  and projecting into basis function $\phi_i(\mathbf{k})$ which gives,
	\begin{equation}
	\Xi^{i}_{\lambda\lambda'}=-\frac{\int_{{FS}_{\lambda}} \frac{dk}{v_F^{\lambda}(\mathbf{k})}\int_{{FS}_{\lambda'}} \frac{dk'}{v_F^{\lambda'}(\mathbf{k}')} \phi_{i}(\mathbf k)V_{\lambda,\lambda'}^s(\mathbf{k},\mathbf{k}')\phi_{i}(\mathbf k')}{\int_{{FS}_{\lambda}} \frac{dk}{v_F^{\lambda}(\mathbf{k})}\phi^2_{i}(\mathbf k)}.
	\label{eigenvalueproblem}
	\end{equation}
	The eigenvector corresponding to the maximum eigenvalue $\xi$ determines the pairing symmetry, where critical temperature is related to $\xi$ as $
	T_c\propto \exp(-1/\xi)$. This relation justifies that maximum positive eigenvalue $\xi$, has a higher critical temperature  and therefore by lowering the temperature the superconducting instability occurs in the corresponding symmetry channel. The point group symmetry of the square lattice allows the following lowest-order even basis functions
		\begin{equation}
\label{bassisschbeyder}
\begin{matrix}
s=1,\\ d_{x^2-y^2}=  (\cos k_x-\cos k_y), \\ 
d_{xy}= \sin k_x\sin k_y,\\ g= (\cos k_x-\cos k_y) \sin k_x\sin k_y\\
g^*=(\cos k_x - \cos k_y)^2 - 4(\sin k_x\sin k_y)^2
\end{matrix}
\end{equation}
that we use to find the maximum eigenvalue $\xi$ and the corresponding  pairing symmetry.

\section{Results}	
\label{results}
In this section we present our results for the susceptibilities and pairing instability of the lattice  model. 
\subsection{Density of states and susceptibilities}	
For the square lattice we use the parameters $t=1$, $t'=0.3t$, and $V_{\rm so}=0.5t$. With this choice for the Rashba spin-orbit coupling, the Fermi contours are largely separated  in momentum space in most fillings. We plot the density of states (DOS) for this set of parameters in Fig. \ref{fighelicity}b. Near the half-filling there are two van Hove singularities due to the spin-split bands, while at $V_{\rm so}=0$ there is only one singularity for each spin species (up and down) as shown in Fig. \ref{fighelicity}c.

Now we turn to the bare susceptibilities. As described in Sec. (\ref{effectivinteraction}), the spin and charge channel in the static limit ($\omega\rightarrow0^+$) are decoupled allowing one to study the instability in each channel separately. That is, we can write $\chi^R=\chi^c\oplus \chi^s$ , where $\chi^c=\chi_{00}$ is the charge susceptibility and $\chi^s=[\chi]_{ij}$ ($i,j=1,2,3$) is the spin susceptibility tensor. For numerical calculation of susceptibilities we mesh grid the Brillouin zone into $100\times100$ $\mathbf k$ points. At each wave vector $\mathbf k$ we evaluate $\chi^c$  and three eigenvalues of the matrix $\chi^s$. We found $\chi^c<0$ in the entire Brillouin zone for all fillings as shown in first row of Fig. \ref{figmeshsucep}. For spin channel we only show  the maximum eigenvalue of $\chi^s$ in second row of Fig. \ref{figmeshsucep}. For the spin channel we found all the eigenvalues are always negative. Therefore, there will be some critical value of $U_c$ ($U_{1c}$) at which the value of det$(1-V X^R)$ vanishes in the spin (charge) channel signaling an spin-density wave (charge-density wave) instability. In Fig. \ref{fighelicity}d we show  the value of the critical Hubbard interaction in different fillings for $U_1=0$. It is clearly seen that the critical $U_c$ is small near the half-filling, where the DOS is large. Note that the values of critical $U_{1c}$ (not shown here) 
generally depends on $U$.

For values of $U$ $(U_{1})$ in the vicinity of the SDW (CDW) critical points, as we will describe in the next subsection, 
the fluctuations of spin (charge) channel play a decisive role in determining the pairing symmetry. However, 
for generic values far away from the critical points, both spin and charge fluctuations conspire to form the pairing 
symmetry which is different from the symmetry expected from individual channels. 
\subsection{Pairing symmetry}
\begin{figure*}[t]
		\includegraphics[width=1\linewidth]{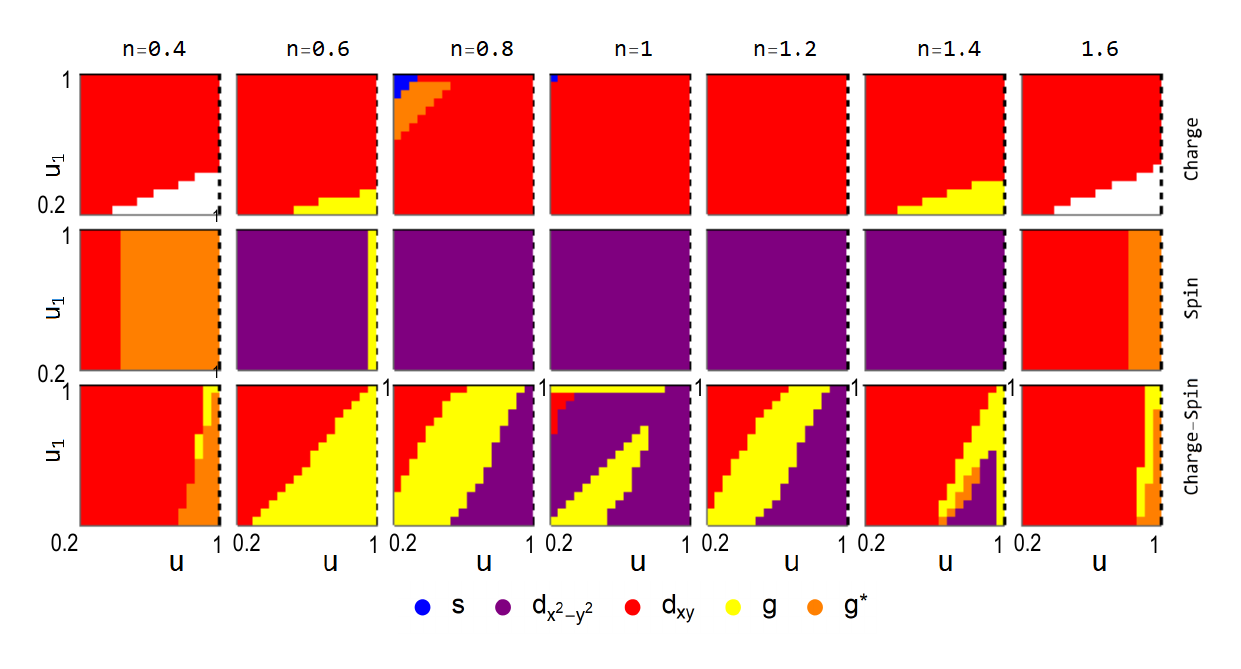}
	\caption[Fermi surface]{(Color online) The phase diagram in the plane of ($u_1=U_1/U_{1c},u=U/U_{c}$). Colors indicate the 
	pairing symmetry for different fillings (right to the left) and for charge-only (first row) spin-only (second row) and charge-spin (third row) channels. In the white region  there is no positive eigenvalue for $s,d_{xy},d_{x^2-y^2}, g$ channels.
	The black thick (dashed) line  at $u_1=1$ ($u=1$) denotes the onset of charge (spin) density wave phase.}
	\label{VPLUSU}
\end{figure*}

We follow the procedure outlined in Sec. (\ref{DOGF}) to determine the pairing symmetry. 
We begin by considering ($u=U/U_c$, $u_1=U_1/U_{1c}$) phase diagram for the filling factors presented in the Fig.~\ref{VPLUSU}. 
We take the effective interaction into account first in separate charge-only (top panels) and spin-only (middle panels) channels. Then 
in order to investigate the competition between the fluctuations in the two channels, we
contrast those separate channels with the situation where both spin and charge channels are taken into account (bottom panels). 
As a general rule, close to $U_{c}$ ($U_{1c}$) boundary the dominant role is played by the spin (charge) fluctuations. 
Further increasing of $U$ ($U_{1}$) beyond $U_c$ ($U_{1c}$) derive the system to the spin (charge) density wave and the ground state will be ordered. Therefore we focus on the square region where $u,u_1 < 1$ and look for superconducting instability in the disordered metallic phase. 

The relevant pairings in the plane of ($u$,$u_1$) at different filling factors are also indicated in Fig.~\ref{VPLUSU}. Different colors stand for pairing symmetry indicated below the panels. In the white region no positive eigenvalue has been found for
the angular momenta $\ell=0,2,4$ (note that $\Delta_\pm(\bf k)$ is an even function of $\bf k$). 
As we can see from the first row of the Fig.~\ref{VPLUSU}, when we consider the charge fluctuations only, 
the phase diagram is dominated by the $d_{xy}$ pairing at and around half-filling. By heavily doping away 
from half-filling, at $n=0.6,1.4$, a $g$-wave pairing for small $u_1$ and large enough Hubbard $u$ appears.  
By further doping away from half-filling, no solution up to angular momentum $\ell=4$ is found which is by the white color. This may correspond to possible higher angular momentum pairing. 

There is a small blue region for small $u$ and large $u_1\lesssim 1$ corresponding to s-wave profile of $\Delta_\pm({\bf k})$. 
The blue (s-wave) and white regions are artifact of overemphasizing the charge fluctuations.
To see this, let us focus on the second row of Fig.~\ref{VPLUSU} which takes only spin fluctuations
into account. As can be seen at and around the half-filling, the dominant pairing is $d_{x^2-y^2}$ (purple region). 
Doping further away from half-filling by either holes or electrons stabilizes the $d_{xy}$ pairing (red region) for
values of $u$ far below the SDW instability. By approaching the SDW instability, a higher angular momentum, $g$-family pairing 
kicks in. For $n=0.6$, still the dominant pairing is $d_{x^2-y^2}$ which eventually gives way to $g$-wave pairing 
by approaching the SDW instability $u\lesssim 1$. For lower electron (hole) density $n=0.4~(1.6)$, the 
phase diagram is divided between the $d_{xy}$ and $g^*$ pairing. The division is almost independent of $u_1$ as
in the spin-only channel, $u_1$ does not play any role. Note that in the charge-only channel, both $u$ and $u_1$ affect
the phase diagram. That is why in the first row, the phase boundaries are not horizontal (i.e. $u$-independent). 

Now let us focus on the third row of Fig.~\ref{VPLUSU} where we let both spin and charge fluctuations to renormalize the
interaction at RPA level. At  and around  half-filling, when $u_1$ is not large, the purple region conquers particularly larger $u$ region.
That is why in this region the red region (due to charge fluctuations) is completely washed out. This 
can be understood in terms of enhancement of spin fluctuations as one approaches SDW critical point. 
Quite generally in the Hubbard model, at half-filling and large enough $u$, the charge
degrees of freedom tend to be frozen and the dominant low-energy fluctuations are those of spin degrees of freedom. 
At half-filling the charge fluctuations find a chance to stabilize a small region corresponding to $u_1\lesssim 1$ and Hubbard $u$ is small. 
Slight deviation from half-filling expands the red region. 

When the strength of $u_1$ and $u$ are comparable, the fluctuations
in both charge (red region) and spin (purple region) will have comparable strength in such a way that they both loose
and give way to $g$-wave pairing indicated by yellow region. This is because the basis function corresponding
to the second largest eigenvalue of Eq.~\eqref{eigenvalueproblem} is generically  dominated by $g$-wave (yellow) pairing. 
That is how the yellow region can take over once the spin (purple) and charge (red) fluctuations can not favor a $d$ wave pairing. By doping away from half-filling with either electrons or holes, the red region arising 
from charge fluctuations expands. The expansion of the charge fluctuations dominated region starts from smaller $u$ when the system
is close to half-filling, and eventually occupies larger $u$ region when $u_1$ is strong enough. 

When the doping crosses the quarter-filling, a lot of phase-space for the charge fluctuations will be 
created. That is why in both second and third row, we obtain qualitatively similar phase diagram
where the major competition is taking place between the red (charge fluctuation dominated phase) and
yellow ($g$-wave pairing).

\begin{figure}[t]
		\includegraphics[width=1\linewidth]{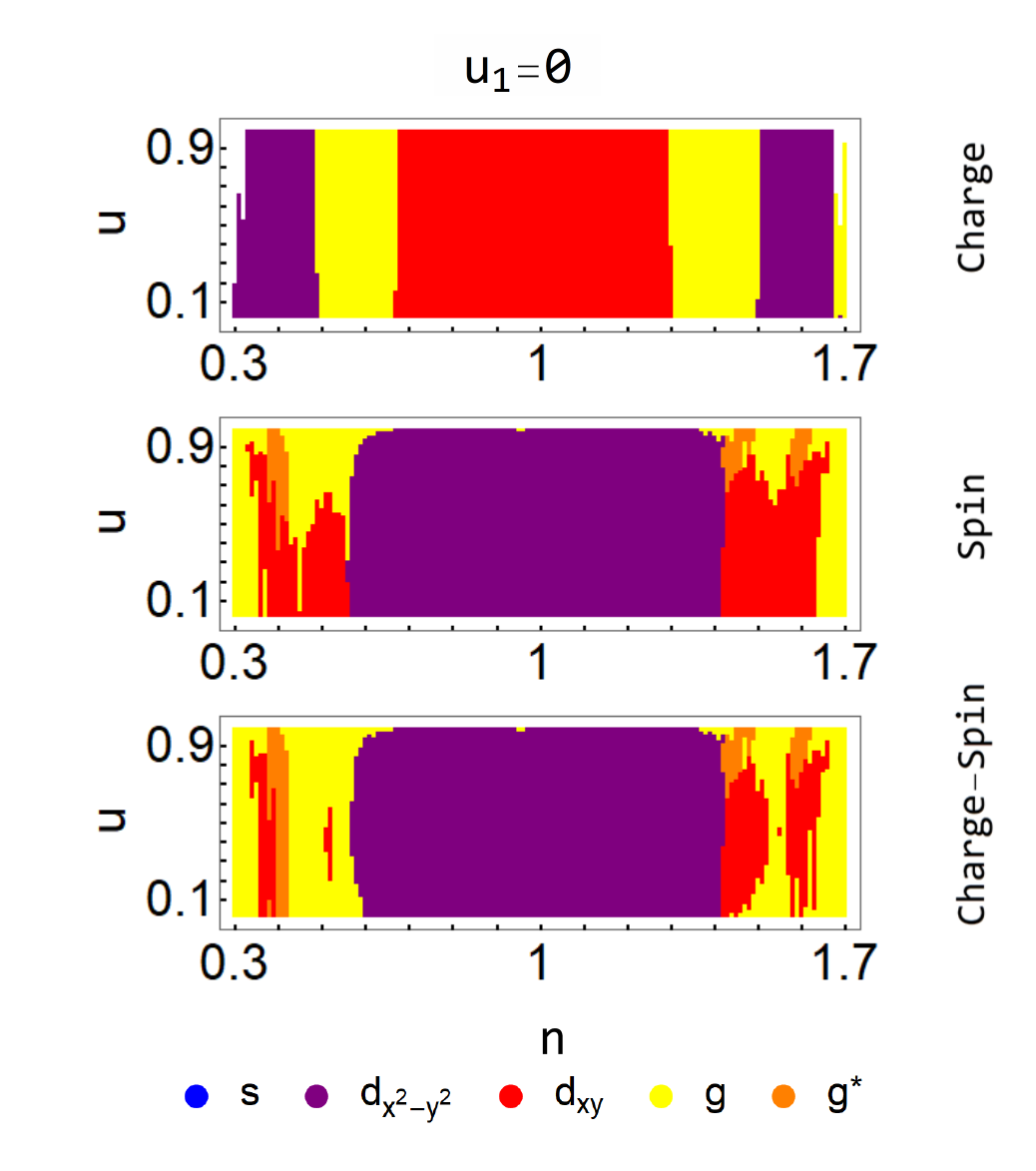}
	\caption{(Color online)  The ($u$,$n$) phase diagram of pairing symmetry for $u_1=0$ are plotted for charge (first row), spin (second row) 
	and charge-spin channel (third row).}
	\label{figsquare1dahomcgg}
\end{figure}

\begin{figure}[t]
		\includegraphics[width=1\linewidth]{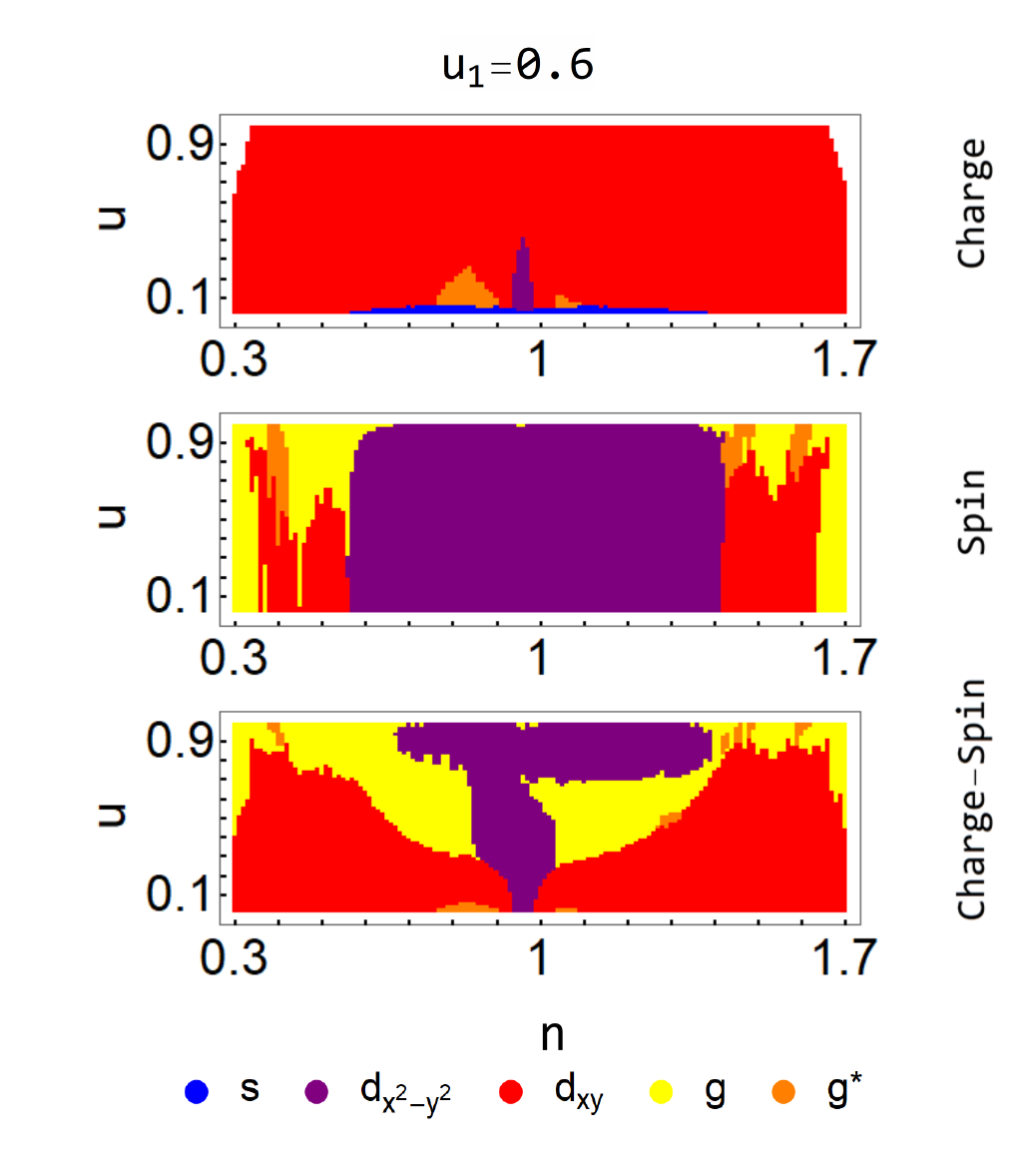}
	\caption{(Color online)  The ($u$,$n$) phase diagram of pairing symmetry for $u_1=0.6$ are plotted for charge (first row), spin (second row) 
	and charge-spin channel (third row).}
	\label{figsquare2dahomcgg}
\end{figure}

To focus on the dependence of the pairing symmetry on electron density, in 
Fig.~\ref{figsquare1dahomcgg}~(\ref{figsquare2dahomcgg}) we plot 
the $u$-$n$ phase diagram for fixed nearest neighbor value of interaction $u_1=0~(0.6)$.
Let us begin with the $u_1=0$ case shown in Fig~\ref{figsquare1dahomcgg}. In this figure, 
we present the pairing symmetry resulting from the fluctuations in charge only, spin only and 
charge-spin together in the first, second and third rows, respectively. 
As can be seen by comparison of the second and third rows, for $u_1=0$ case,
the purple area surrounding the half-filling and region around it is quite
similar in both cases. This means that
the spin channel plays the dominant role in determining the pairing symmetry. 
By moving to low carrier density (in either electron or hole sides), the
$d_{x^2-y^2}$ pairing looses, and the main competition will be between $d_{xy}$
and $g$-wave family pairing. This family consists in the standard $g$-wave (yellow) and
and $g^*$-wave pairing (orange). 
In these regimes, when $u$ is large enough, again the
second and third rows in Fig.~\ref{figsquare1dahomcgg} are similar, which is 
natural, as the spin fluctuations are the most strong for $u\lesssim 1$.
Upon lowering $u$, the third row phase diagram starts to deviate from the
second row, and the $g$-wave and $g^*$-wave pairing will win. Again as discussed before, 
when one plots a similar phase diagram with the second largest eigenvalue, 
the major parts of the phase diagram, and in particular the low-carrier density
regime turns out to be $g$-family (yellow/orange) dominated. Hence the $g$-family pairing
sets in, in the case of comparable strength between the purple and
red regions. In such situation both purple and red loose, and yellow/orange region 
takes over. This can be clearly seen by comparison of the first and second
rows in Fig.~\ref{figsquare1dahomcgg}.

Similarly, in Fig.~\ref{figsquare2dahomcgg} we have plotted the phase diagram in the $(u,n)$ plane for
the fixed value of $u_1=0.6$. Again in first (second) row we have only considered the charge (spin) 
fluctuations, while in the third row we have considered spin and charge fluctuations together. 
The difference between this figure, and Fig.~\ref{figsquare1dahomcgg} is the value of $u_1$. 
As can be seen from Eq.~\eqref{effinteraction}, the nearest neighbor interaction $u_1$ affects 
only the charge component of the effective interaction. That is why the second row in both 
Figs.~\ref{figsquare1dahomcgg} and~\ref{figsquare2dahomcgg} are identical. However, 
the first row in these two figures are drastically different. As a result of overemphasizing
the charge fluctuations (by choosing to focus only on the charge channel), major parts of 
the $(u,n)$ phase diagram is dominated by $d_{xy}$ (red region) pairing. Small region of
$s$-wave (blue) pairing also appears. Such a blue region is entirely absent in the second and third
rows. This is actually artifact of limiting the total energy minimization to few lowest
angular momentum basis functions. Indeed, allowing for higher angular momentum basis 
such as $g^*$, they will become dominant over the $s$-wave pairing.
Therefore the $s$-wave pairing is artifact of limited number of basis functions,
and can be removed by including more and more basis functions. 
After all, in second and third rows of Fig.~\ref{figsquare2dahomcgg} there are
no $s$-wave pairing which simply means that the spin fluctuations do not favor $s$-wave pairing. 

It is instructive to compare third rows of Figs.~\ref{figsquare1dahomcgg} and~\ref{figsquare2dahomcgg}.
As can be seen, for large $u\lesssim 1$ in both $u_1=0$ and $u_1=0.6$ cases,  the spin fluctuations play the dominant
role, and the resulting $d_{x^2-y^2}$ pairing (purple) region conquers the half-filling and region around it. 
By reducing $u$ to smaller values, the situation in $u_1=0$ case of Fig.~\ref{figsquare1dahomcgg} does not
change much, while in Fig.~\ref{figsquare2dahomcgg} first a $g$-wave (yellow) pairing kicks in. Then by 
further reduction in $u$, the $d_{xy}$-wave pairing dominates. For slightly hole doped case,
the purple region continues to win. Needless to say, the drastic difference between the second
and third rows in Fig.~\ref{figsquare2dahomcgg} signifies the importance of the charge fluctuations. When they are included, they drastically change the picture arising from spin-only fluctuations.


\subsection{Degeneracy of the solutions and possible topological superconductivity}
So far we have determined the relevant pairing function for different filling and interaction parameters $u,u_1$. 
However we have not yet considered the degeneracy of the solution. The relation between the degeneracy of the
eigenvalues of Eq.~\eqref{eigenvalueproblem} and topological superconductivity is as follows:
Suppose the two largest positive eigenvalues of Eq.~\eqref{eigenvalueproblem} are $\xi_0,\xi_1$. 
Their relative difference can be quantified by $\delta \xi=(\xi_0-\xi_1)/\xi_0$. $\delta\xi\to 0$ indicates nearly degenerate solutions. In Ref. [\onlinecite{PhysRevB.81.024504}] it is shown that the degenerate pairings belonging to two-dimensional irreducible representation can spontaneously break the time-reversal symmetry and a pairing with nontrivial winding number develops in the system. Even if the pairing symmetries do not belong to higher-dimensional representations, but with the same angular momentum, upon lowering the temperature a phase transition to a complex pairing state occurs, e.g. in UPt$_{3}$~\cite{Schemm190} .      
For nearly degenerate channels $d_{xy}$ and $d_{x^2-y^2}$ a complex $d \pm id$ combination
is favored as it avoids the nodes~\cite{Gonge1602579}. The latter state has been realized in epitaxial Bi/Ni bilayer system~\cite{Gonge1602579}. Such a combination gives rise to non-trivial topology
in the form of a  non-zero winding number~\cite{samokhin2015symmetry}. Thus generically when two solutions with 
the same angular momentum are degenerate, a time-reversal symmetry breaking chiral superconducting order can be established. 
We anticipate the degeneracies to happen at the phase boundaries between superconducting orders with different symmetries. Of particular interest is the degenerate boundaries between $d_{xy}$ (red) and $d_{x^2-y^2}$ (purple) near the half-filing in Fig.~\ref{figsquare2dahomcgg}. Therefore we expect a superconducting state with non-zero winding number $\ell=\pm2$~\cite{Gonge1602579}. We have to emphasize that such superconducting instability may change the phase diagram near the boundaries, but the exact determination of the phase diagram is beyond the scope of current study, and we leave it for future study.    



 \subsection{Gap structure in original spin basis}
 	\begin{figure}
		\includegraphics[width=1\linewidth]{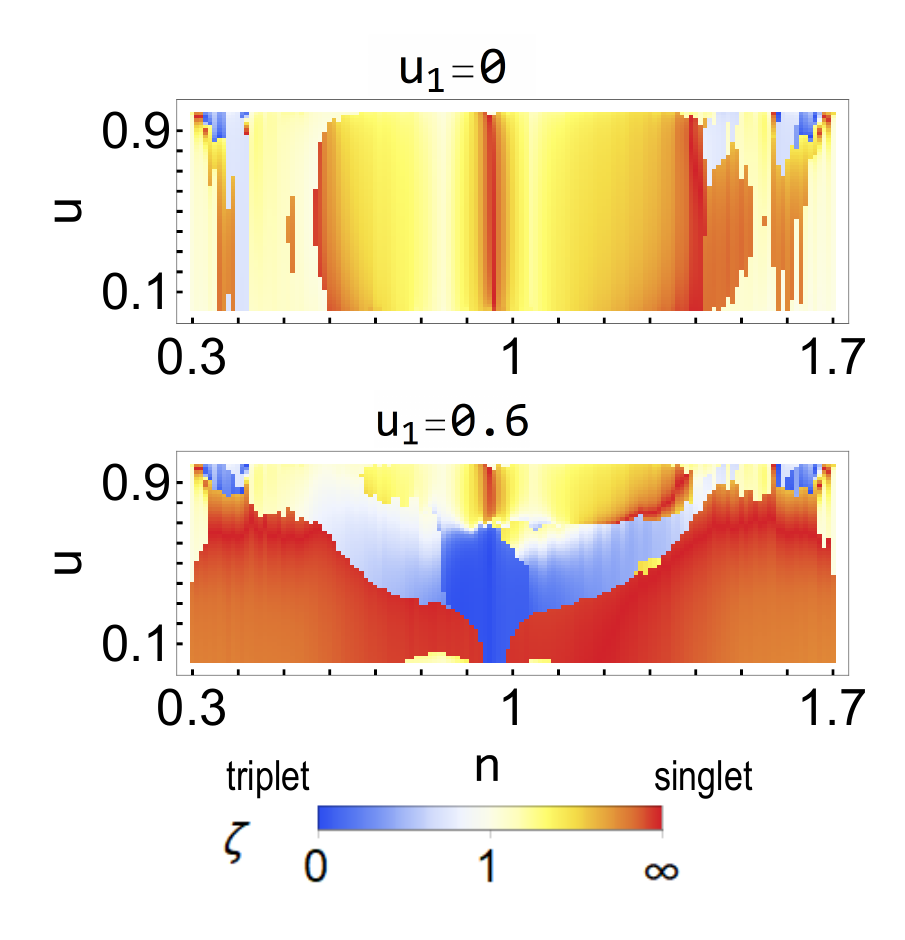}
 	\caption[singlet and triplet]{(Color online) The $(u,n)$ phase diagram for the ratio of triplet and singlet component of gap function $\zeta$ are plotted for $u_1=0$ (top panel) and $u_1=0.6$ (bottom panel).}
 	\label{singlettriplet}
 \end{figure}
In general, in the presence of Rashba spin-orbit coupling and inversion symmetry breaking, the pairing wave function is a mixture of spin singlet and triplet components.\cite{gor2001superconducting} Therefore, the total spin is not a good quantum number to label the pairing wave functions, nor is parity. However, for a multi-component superconductor arising from different Fermi contours, the superconducting wave function can have purely singlet or triplet character as described below. We rewrite the pairing Hamiltonian (\ref{HIMF}) in terms of original spin degrees of freedom as\cite{PhysRevB.76.094516}
\begin{equation}
H_{MF}=\sum_{{\mathbf{k}}\tau\tau'} \left(\Delta_{\tau\tau'}(\mathbf{k}) c^\dagger_{{\mathbf{k}},\tau} c^\dagger_{-{\mathbf{k}},\tau'}+h.c.\right),
\end{equation}
where,
\begin{equation} 
\Delta_{\tau\tau'}(\mathbf{k}) =\sum_{\lambda}\Delta_{\lambda}(\mathbf{k})(-1)^{\frac{\tau+1}{2}} ( t_{\lambda\mathbf{k}})^{\frac{\tau+\tau'}{2}}.
\end{equation}
The above equation can be written explicitly in terms of singlet and triplet components as
\begin{equation}
\Delta_{\tau\tau'}(\mathbf{k}) = \psi_{\mathbf k}(i\sigma_y)_{\tau\tau'}+\mathbf d_{\mathbf k}\cdot(i\sigma_y\boldsymbol\sigma)_{\tau\tau'},
\end{equation}
where 
\begin{eqnarray}
\label{singlet} \psi_{\mathbf{k}}=-\frac{\Delta_{+}(\mathbf{k})+\Delta_{-}(\mathbf{k})}{2}
\end{eqnarray}	 
is the singlet amplitude and
\begin{equation}
\mathbf d_{\mathbf k}=\frac{\Delta_{+}(\mathbf{k})-\Delta_{-}(\mathbf{k})}{2}\mathbf{\hat{g}}_{\mathbf k}
\label{triplet}
\end{equation}	 
is the triplet one. Note that $\psi_{\mathbf{k}}=\psi_{-\mathbf{k}}$ is an even function and $d_{\mathbf k}=-d_{-\mathbf k} $ is an odd one, since $\Delta_{\lambda}(\mathbf{k})=\Delta_{\lambda}(-\mathbf{k})$ and $\mathbf{\hat{g}_{\mathbf k}}=-\mathbf{\hat{g}}_{-\mathbf k}$. Therefore if the pairing symmetry on both Fermi contours are the same, i.e.  $\Delta_{\lambda}(\mathbf k)=\Delta_{\lambda} \phi(\mathbf k)$, at special phases where $\Delta_{+}=-\Delta_{-} $ ($\Delta_{+}=\Delta_{-} $) the singlet (triplet) component vanishes and consequently the pairing become purely triplet (singlet). 
In order to compare the triplet and singlet  component of the gap functions we define $\zeta=\left|(\Delta_{+}+\Delta_{-})/(\Delta_{+}-\Delta_{-})\right|$. Obviously $\zeta\to 0$ indicates the dominance of the triplet pairing while the opposite limit $\zeta \to \infty$
corresponds to the singlet pairing. 

In Fig.~\ref{singlettriplet}, the top and bottom panels, respectively, show  the values of $\zeta$ for $u_1=0$ and $u_1=0.6$.  
As we can see in the case of $u_1=0$ the phase diagram is dominated by $\zeta>1$, and therefore the pairing is
more inclined towards the singlet character. Note that due to Rashba spin-orbit coupling, it can not be a pure singlet which only happens when $\zeta\to\infty$. 
There is also a small light blue ($\zeta<1$, triplet dominated) region in Fig.~\ref{singlettriplet} which corresponds to the 
the $g$-family region of Fig.~\ref{figsquare1dahomcgg}. Note however that the $g$-wave pairing also appears in 
light yellow regions which means that far away from half-filling their singlet character can become slightly 
stronger than the triplet. 

Now let us discuss the $u_1=0.6$ panel of Fig.~\ref{singlettriplet}. 
The large $u\lesssim 1$ part of this panel is generally similar to the $u_1=0$ panel. 
However, it turns out that  the singlet (triplet) component become sharper as indicated by the colors intensity.  
Comparison with third row of Fig.~\ref{figsquare2dahomcgg} shows that the singlet dominated region
in the $u_1=0.6$ panel of Fig.~\ref{singlettriplet} corresponds to $d_{xy}$ pairing. 
Around $n\approx 0.95$ there is strong triplet component (darker blue) which corresponds to
$d_{x^2-y^2}$ region. This region extends over a larger region by increasing $u$ but with lower color intensity. Therefor the nearest-neighbor interaction can stabilize the triplet pairing near the half filling.   

Recently Greco and Schneyder in Ref.~\onlinecite{greco2017topological} have found  the triplet solution 
for $u_1=0$ and chemical potentials that lie between two van Hove singularities. However as seen from
the upper panel of Fig.~\ref{singlettriplet} corresponding to $u_1=0$, we find singlet-dominated pairing near the half filling 
which takes place in the $d_{x^2-y^2}$ channel. There can be two possible reasons for this discrepancy. First, we worked out the pairing symmetry in the band basis which are appropriate basis in the presence of the strong spin-orbit coupling. The second and perhaps more important reason is that in computation of the effective interaction, rather than limiting ourselves to transverse or longitudinal portions of the susceptibility matrix, we have considered the full tensorial structure of the susceptibility in the spin-charge basis. 
Comparing our results with Ref.~\onlinecite{doi:10.7566/JPSJ.82.014702}, we see that the phase diagram presented in 
Fig.~\ref{VPLUSU} for filling $n=0.8$ is qualitatively similar to the results presented in this reference.


\section{Conclusions \label{conclusions}}
In conclusion, we have studied the effect of the large Rashba spin-orbit interaction on the interaction driven superconducting instability 
on the square lattice with on-site (Hubbard $U$) and nearest neighbor interaction ($U_1$). We developed a complete RPA effective interaction by taking into account the full
tensorial structure of the susceptibility in the spin-charge channels, rather than picking the singlet or triplet channels only. We focused on a range of interactions where the system is metallic with no magnetic and/or charge orderings. We mapped out superconducting phase diagrams in the parameter space spanned by interactions and fillings. In the absence of the nearest neighbor Coulomb interaction $U_1$, generically the dominant pairing is in $d_{x^2-y^2}$ channel, and the pairing interaction mainly arises from the spin-fluctuations. The nearest-neighbor interaction, however, increases the charge fluctuations and favors  the $d_{xy}$ symmetry at small $u$. In the regime where both interactions are comparable spin and charge fluctuations are strong and higher angular momentum pairing states are favored. We also pointed out by evaluating the degenerate solutions near the phase boundaries the system can possibly break the time-reversal symmetry spontaneously and a topological superconductor can take over. We also showed that the nearest-neighbor interaction can stabilize a triplet pairing symmetry near the half filling, where the amplitude of the singlet component almost vanishes.

\section{Acknowledgements}
M. K. acknowledges the support from the Sharif University
of Technology under Grant No. G690208.
S. A. J. acknowledges Sharif Univ. of Tech. and Science Elites Federation of Iran.

	
	\appendix
	\section{Interaction matrix}
	\label{Interactionmatrix}
	We can rewrite interaction matrix (\ref{hubbard}) in terms of density operators (\ref{density}),
	\begin{equation}
	\label{UTOU}
		 n_{i\uparrow}n_{i\downarrow}=-\frac{1}{8}\sum_{\alpha=0}^3(-1)^{\delta_{\alpha,0}}\left( \left(\begin{matrix}
		c^\dagger_{i\uparrow}\\		c^\dagger_{i\downarrow}
		\end{matrix}\right)^T\sigma^{\alpha}\left(\begin{matrix}
		c_{i\uparrow}\\		c_{i\downarrow}
		\end{matrix}\right)\right)^2+\frac{1}{4}\sum_{\tau}n_{i\tau},
	\end{equation}
	where $\delta_{\alpha,0}=1$ if $\alpha=0$, otherwise vanishes. 
	By Fourier transformation and replacing $U/8\rightarrow U$ with adding the last term of Eq. (\ref{UTOU}) to the chemical potential we can rewrite Eq. (\ref{hubbard}) as follows,
	\begin{eqnarray}
\label{hubbard2}
H_I=&\frac{1}{N}\sum_{q}\left[U+U_1(\cos q_x+\cos q_y) \right]\rho_{0,\mathbf{q}}\rho_{0,-\mathbf{q}}\nonumber\\
&-\frac{U}{N}\sum_{q,i=1}^3\rho_{i,\mathbf{q}}\rho_{i,-\mathbf{q}} .
\end{eqnarray}
Consequently, we can rewrite $H_I$ in the compact form presented in Eq. (\ref{HI}).\\

\section{Derivation of the effective interaction}
The effective interaction can be decomposed into charge and spin channels. The matrix elements $V_{\alpha,\beta}$ of the interaction
are written in the basis of charge ($\alpha=0$) and spin ($\alpha=1,2,3$). Therefore $\rho_0$ will be the charge density, while $\rho_{i}$
with $i=1,2,3$ corresponds to three components of the spin density. 
The RPA effective interaction reads as follows 
\begin{equation}
-\hat{V}^{\rm eff}(\omega=0,\mathbf{q})=-\hat{V}(\mathbf{q})+\hat{V}(\mathbf{q})[-\hat{\chi}(\omega=0,\mathbf q)]\hat{V}(\mathbf{q})+\dots,
\end{equation}
which can be written in a compact form as
\begin{equation}
	\hat{V}^{\rm eff}(\omega=0,\mathbf q)=\left[1+\hat{V}(\mathbf q)\hat{\chi}(\omega=0,\mathbf q)] \right)^{-1} \hat{V}(\mathbf q).
\end{equation}

\section{Vanishing of the spin-charge cross term in the static susceptibility matrix}
\label{vanishing}
By using the definition of the form factor (\ref{formfactor}) it follows that~\cite{pletyukhov2006screening,pletyukhov2007charge}
\begin{eqnarray}
\label{eq1}
	F^{{\beta,\alpha}*}_{\mathbf k,\mathbf k+\mathbf q;\lambda,\lambda'}=	F^{{\alpha,\beta}}_{\mathbf k,\mathbf k+\mathbf q;\lambda,\lambda'}\\
	\label{eq2}
F^{{\alpha,\beta}}_{\mathbf k,\mathbf k+\mathbf q;\lambda,\lambda'}=s^{\alpha,\beta}	F^{{\alpha,\beta}}_{-\mathbf k-\mathbf q,-\mathbf k;\lambda',\lambda}
\end{eqnarray}
where $s^{\alpha,\beta}=-1$ if either $\alpha=\{0\},\beta=\{1,2,3\}$ or $\alpha=\{1,2,3\},\beta=\{0\}$. Otherwise it  equals to one.  
By changing integration variable $k\rightarrow -k-q$ and using Eq.~(\ref{eq2}) we can rewrite  Eq.~(\ref{XR}) as
\begin{widetext}
			\begin{equation}
	\label{XR2}
	[\chi^R(\omega,\mathbf q)]^{\alpha,\beta}=s^{\alpha,\beta}\int\frac{d^2k}{4\pi^2}\sum_{\lambda \lambda'=\pm 1}  \frac{n_F(\varepsilon_{\lambda\mathbf k})-n_F(\varepsilon_{\lambda'\mathbf k+\mathbf q})}{-\omega-i0^++\varepsilon_{\lambda\mathbf k}-\varepsilon_{\lambda'\mathbf k+\mathbf q}}[F_{\mathbf k,\mathbf k+\mathbf q;\lambda,\lambda'}]^{\alpha,\beta}.
	\end{equation}
However, for $\chi^{\dagger}$ by help of  Eq. (\ref{eq2}) we can write
	\begin{equation}
	\label{XR3}
	[\chi^R(\omega,\mathbf q)]^{\beta,\alpha*}=\int\frac{d^2k}{4\pi^2}\sum_{\lambda \lambda'=\pm 1}  \frac{n_F(\varepsilon_{\lambda\mathbf k})-n_F(\varepsilon_{\lambda'\mathbf k+\mathbf q})}{\omega-i0^++\varepsilon_{\lambda\mathbf k}-\varepsilon_{\lambda'\mathbf k+\mathbf q}}[F_{\mathbf k,\mathbf k+\mathbf q;\lambda,\lambda'}]^{\alpha,\beta}.
	\end{equation}
\end{widetext}
The hermiticity of the effective interaction requires $\chi^{\dagger}(\omega=0,\mathbf q)=\chi(\omega=0,\mathbf q)$.
Therefore, Eqs.~(\ref{XR2}) and~(\ref{XR3}) force the cross terms of the $\hat{\chi}^R(\omega=0,\mathbf q)$ between charge and spin sectors vanish.

%

\end{document}